
\input manumac.tex
\def\M{\cal M}

\twelvepoint

\def\dplus{=\hskip-5pt \raise 0.7pt\hbox{${}_\vert$} ^{\phantom 7}}
\def\dplusup{=\hskip-5.1pt \raise 5.4pt\hbox{${}_\vert$} ^{\phantom 7}}

\font\bb=msbm10 scaled 1200

\font\twelvebb=msbm10 scaled 1200
\font\ninebb=msbm10 scaled 900
\font\sevenbb=msbm10 scaled 700
\newfam\bbfam
\def\bb{\fam\bbfam\twelvebb}
\textfont\bbfam=\twelvebb
\scriptfont\bbfam=\ninebb
\scriptscriptfont\bbfam=\sevenbb

\def\M{{\cal M}}
\def\com{{\bb C}}

\font\twelvebbb=msbm5 scaled 1200
\font\ninebbb=msbm5 scaled 900
\font\sevenbbb=msbm5 scaled 700
\newfam\bbbfam

\textfont\bbbfam=\twelvebbb
\scriptfont\bbbfam=\ninebbb
\scriptscriptfont\bbbfam=\sevenbbb

\font\bbbb=msbm7 scaled 1200

\font\twelvebbbb=msbm7 scaled 1200
\font\ninebbbb=msbm7 scaled 900
\font\sevenbbbb=msbm7 scaled 700
\newfam\bbbbfam
\def\bbbb{\fam\bbbbfam\twelvebbbb}
\textfont\bbbbfam=\twelvebbbb
\scriptfont\bbbbfam=\ninebbbb
\scriptscriptfont\bbbbfam=\sevenbbbb

\line{\hfil DAMTP/R-93/17}
\line{\hfil KCL-93-8}
\line{\hfil July 1993}

\vskip 0.4truecm
\centerline{\bf MASSIVE SIGMA MODELS WITH (p,q) SUPERSYMMETRY}
\bigskip
\centerline{\bf G. Papadopoulos}
\medskip
\centerline{Dept. of Mathematics, King's College London,
England\footnote{$^*$}{Address after Oct. 1 1993: II Institute for theoretical
physics, Univ. of Hamburg, Germany.} }
\bigskip
\centerline{\it and}
\bigskip
\centerline{\bf P.K. Townsend}
\medskip
\centerline{DAMTP, University of Cambridge, Cambridge, England}

\vskip 1cm
\AB
We determine the general scalar potential consistent with (p,q) supersymmetry
in two-dimensional non-linear sigma models with torsion, generalizing previous
results for special cases. We thereby find many new supersymmetric sigma models
with potentials, including new (2,2) and (4,4) models.
\AE

\vfill\eject
\noindent {\bf 1. Introduction }
\medskip

The general (p,q) supersymmetry algebra for two-dimensional (d=2) Minkowski
space field theories is spanned by $p$ Hermitian spinorial charges $\{ Q_+^I
;I=1,\dots ,p\}$ of one chirality, $q$ charges $\{ Q_-^{I'} ;I'=1,\dots ,q\}$
of
the other chirality, and the self-dual and anti-self-dual components, $P_\dplus
,P_=$, of the 2-momentum (we use a Lorentz charge notation for d=2 spinors and
vectors). Excluding spinorial or tensorial central charges, there remains only
the possibility of an additional $pq$ scalar central charges, $Z^{II'}$. The
non-zero (anti)commutation relations of this algebra are
$$
\{Q_+^I, Q_+^J\}=2\delta^{IJ}P_\dplus \qquad
\{Q_-^{I'},Q_-^{J'}\}=2\delta^{I'J'}P_=
\qquad \{Q_+^I,Q_-^{I'}\}= Z^{II'}\ .
\eqno (1.1)
$$
All d=2 supersymmetric field theories have at least (1,0) or (0,1)
supersymmetry
and, since (0,1) is the parity reflection of (1,0), we may assume without loss
of generality that all d=2 supersymmetric field theories are (1,0)
supersymmetric. If we restrict our attention to scalar and spinor fields only
then all such theories can be written in terms of the (1,0) scalar superfields
$\{\phi^i(x,\theta^+); i=1,\dots,D\}$, which define a map
$\phi$ from (1,0) superspace, $\Sigma^{(1,0)}$, into a $D$-dimensional
Riemannian target manifold ${\cal M}$, and the spinor superfields
$\{\psi_-^a(x,\theta ^+); a=1,\dots,n\}$, which define a section $\psi$ of the
vector bundle $S_{-}\otimes \phi^*\xi$ where $\xi$ is a vector bundle over
${\cal M}$ of rank $n$ and $S_-$ is the spin bundle over $\Sigma^{(1,0)}$.
Subject to the further restriction that all scalar field equations be second
order, the action for the general (p,q)-supersymmetric model, generally called
a `sigma-model', can be written as the (1,0)-superspace integral [1]
$$
S=\int\! d^2 x d \theta ^+ \big\{ D_+\phi^i\partial_=\phi^j (g_{ij} +b_{ij})
+i\psi_-^a\nabla_+\psi_-^b \; h_{ab} +ims_a\psi_-^a\big\} \ ,
\eqno (1.2)
$$
where $m$ is a constant with dimensions of mass, $D_+ =
i\partial /\partial\theta^+ +\theta^+\partial_\dplus$ is the (real)
supercovariant derivative satisfying $D_+^2=i\partial_\dplus$, and  $$
\nabla_+\psi_-^b\equiv (D_+\psi_-^b + D_+\phi^i \Omega_{i}{}^b{}_c\psi_-^c)\ .
\eqno (1.3)
$$
Thus, in this formulation, the general (p,q)-supersymmetric model is
characterised by (i) a metric $g$ on $\M$, which we require to be positive
definite in order that the Hamiltonian be positive semi-definite (ii) a
two-form $b$ on $\M$; this need be defined only locally since it is $H=db$ that
occurs in the field equations (as a torsion tensor), (iii) a metric $h$ and
connection $\Omega$ on the fibre $\xi$; without loss of generality we can
choose $h$ to be covariantly constant, i.e.
$$
\nabla_i h_{ab}=0\ ,
\eqno (1.4)
$$
and (iv), when $m\ne0$, a section $s$ of $\xi$.

When $m=0$ this action is classically conformally invariant. It will be
shown below that when $m\ne0$ the component action contains the potential term
$$
V(\phi) ={1\over4}m^2 h^{ab}s_a(\phi)s_b(\phi)\ .
\eqno (1.5)
$$
In many models of interest the section $s$, and hence the potential $V$, will
have isolated zeros. Linearisation about an isolated zero of the potential
yields a massive supersymmetric field theory with mass proportional to $m$. We
shall therefore refer to the general model with $m\ne 0$ as a `massive
sigma-model'. Note that when $m\ne 0$, the fibre metric $h$ (as well as the
target space metric $g$) must be positive definite for the Hamiltonian to be
positive semi-definite. We henceforth assume that $h$ is positive definite, in
which case the structure group of the bundle $\xi$ is a subgroup of $O(n)$.

As written, the action (1.2) has {\sl manifest} (1,0) supersymmetry. If certain
conditions on the couplings $g$, $b$, $h$, $\Omega$, and $s$ are satisfied it
will have further, although non-manifest, supersymmetries. The
conditions for (p,q) supersymmetry in the massless case, $m=0$, have been
thoroughly investigated [2,3,4] in past years because of the applications of
these sigma-models to string theory and conformal field theory. More recently
attention has focused on certain massive models. In particular it has
been shown that some massive (2,2) supersymmetric sigma models are
integrable quantum field theories (see e.g. [5]). A feature of these models is
that they admit solitons which interpolate between distinct zeros of the
potential and carry a complex topological charge which appears in the
(on-shell)
supersymmetry algebra as a central charge. More recently, it has been shown
that there exist (4,4)-supersymmetric models with solitons that carry a
quaternionic charge which again appears in the supersymmetry algebra as a
central charge [6]. Also, certain (2,0) massive models have acquired importance
in the context of the Landau-Ginsburg approach to integrable models [7].

The principal purpose of this paper is to provide a {\sl complete} analysis of
the conditions required for (p,q) supersymmetry in massive sigma-models.
A study of the conditions for (p,p) supersymmetry for models without torsion
was
undertaken some ten years ago [8]. Very recently, results for massive models
{\sl with} torsion, i.e. with $H\ne0$, were presented for off-shell (p,0) and
(1,1) models [1] using (1,0) and (p,0) superfield methods.
Here we use the (1,0) superfield methods to determine the conditions
required for on-shell (p,0) supersymmetry. All remaining models can be
considered as special cases of the general (1,1) model. In this case,
the bundle $\xi$ is isomorphic to the tangent bundle of ${\cal M}$ and the
general form of the section $s$ is $s_i= (u-X)_i$,  where $X$ is any
Killing vector field of the target space $\M$ (should one exist) and $u$ is a
one-form on $\M$ defined by $\iota_X H=du$. The potential $V$ is therefore [1]
$$
V= {m^2\over4}g^{ij}(u-X)_i(u-X)_j\ .
\eqno (1.6)
$$
The Noether charge corresponding to the symmetry generated
by $X$ appears in the (1,1) supersymmetry algebra as a central
charge [8]. For (p,q) supersymmetric models with $p>1$, the supersymmetry
algebra of the massless model has an $SO(p)\times SO(q)$ automorphism group
which acts in the obvious way on the central charge matrix $Z^{II'}$. Since $X$
is associated with a {\sl particular} central charge, the potential
(1.6) appears to break the $SO(p)\times SO(q)$ symmetry. However, as we shall
see, (p,q) supersymmetry implies that the potential $V$ can be written in many
different but equivalent ways. A consequence of this fact is that $V$ is
actually $SO(p)\times SO(q)$ invariant (up to a constant). There are
also further restrictions on $X$ and $u$ imposed by (p,q) supersymmetry which
we
state and analyse.

The (2,2) and (4,4) models are of particular current interest. For these cases
one can use the $SO(p)\times SO(q)$ invariance to diagonalize $Z^{II'}$.
The (p,p) supersymetry then implies that all the diagonal elements are
simply related to the Killing vector field $X$. For the (2,2) models with zero
torsion we find agreement with previous results [8] but we believe that the
(2,2) models with non-zero torsion and non-zero central charge are new. It
would be of interest to determine their soliton solutions and whether any of
them are integrable quantum field theories. For the (4,4) case it was known
that zero torsion massive models exist with $u=0$ and $X$ tri-holomorphic. Here
we prove that $u=0$ and $X$ is tri-holomorphic for all (4,4) models with zero
torsion, but for {\sl non-zero} torsion we find new (off-shell) models for
which $X=0$ but $u\ne0$.

The results presented here will have implications for the renormalization
properties of the quantum theory of massive supersymmetric sigma models. For
example, previous results concerning the ultraviolet finiteness of massless
off-shell supersymmetric (4,q) models [4] can be extended to massive (4,q)
models for which the supersymmetry algebra closes off-shell without central
charges, so some of the models constructed here will be ultraviolet finite.

Finally, we remark that an alternative method of constructing (off-shell)
massive (1,1)-supersymmetric sigma models based on gauged massless sigma models
was also presented in [1]. This method allows an alternative construction of
massive off-shell (p,q)-supersymmetric models [9].

\medskip
\noindent {\bf 2. (1,0) Supersymmetry }
\medskip

We begin with a discussion of some general features of the action (1.2).
To determine the component version of this action we define
the component fields contained in $(\phi^i,\psi_-^a)$ by
$$
\phi^i=\phi^i| \qquad \lambda_+^i =D_+\phi^i | \qquad \psi_-^a = \psi_-^a |
\qquad
F^a=\nabla_+\psi_-^a |
\eqno (2.1)
$$
where the vertical bar indicates the $\theta^+=0$ component of a superfield.
Then (1.2) becomes
$$
\eqalign{
S =\int\! d^2 x\big\{ &\partial_\dplus\phi^i\partial_=\phi^j (g_{ij}+b_{ij})
+ ig_{ij}\lambda_+^i\nabla_=^{(+)}\lambda_+^j - i\psi_-^a\nabla_+\psi_-^b
h_{ab}
 -{1\over2}\psi_-^a\psi_-^b\lambda_+^i\lambda_+^j F_{ijab}\cr &+ F^aF^b h_{ab}
+m\nabla_i s_a \lambda_+^i\psi_-^a + ms_a F^a\big\}\ ,\cr}
\eqno (2.2)
$$
where $\nabla^{(\pm)}_=$ is the covariantization of $\partial_=$ with the
pull-back of the connection with
torsion
$$
\Gamma^{(\pm)}_{ij}{}^k = \big\{\matrix{ k\cr ij\cr}\big\} \pm  H_{ij}{}^k\ ,
\eqno (2.3)
$$
so that
$\nabla_=^{(\pm)}\lambda_+^k= \partial_=\lambda_+^k +
\Gamma^{(\pm)}_{ij}{}^k \partial_= \phi ^i \lambda_+^j $.
The torsion tensor $H_{ijk}$ is given by
$$
H_{ijk}={1\over2}(\partial_ib_{jk} +\partial_k b_{ij} + \partial_j b_{ki})
\equiv
{3\over2}\partial_{[i} b_{jk]}\ ,
\eqno (2.4)
$$
and $F_{ijab}= h_{ac} F_{ij}{}^c{}_b$, where
$$
F_{ij}{}^a{}_b = \partial_i\Omega_j{}^a{}_b -\partial_j\Omega_i{}^a{}_b
+\Omega_i{}^a{}_c\Omega_j{}^c{}_b
-\Omega_j{}^a{}_c \Omega_i{}^c{}_b\ .
\eqno (2.5)
$$
Elimination of $F^a$ from (2.2) by means of its algebraic field equation yields
the scalar potential $V$ of (1.5).

Returning now to the (1,0)-superspace action (1.2), we note that its variation
with respect to the {\sl arbitrary} variations $\delta\phi^i$ and
$\delta\psi_-^a$ of $\phi^i$ and $\psi_-^a$ is (up to a surface
term)
$$
\delta S=\int\! d^2 xd\theta^+\big\{ \delta\phi^i {\cal S}_{-i}
+\Delta\psi_-^a{\cal
S}_a\big\}\  ,
\eqno (2.6)
$$
where
$$
\Delta\psi_-^a\equiv \delta\psi_-^a +\delta\phi^i\psi_-^b\Omega_i{}^a{}_b
\eqno (2.7)
$$
is the covariantization of $\delta\psi_-^a$, and
$$
\eqalign{
{\cal S}_{-i} &\equiv -2g_{ij}\nabla^{(-)}_+\partial_=\phi^j -i\psi_-^a\psi_-^b
D_+\phi^j
F_{ijab} + im\nabla_i s_a\psi_-^a \cr
{\cal S}^a &\equiv 2i\nabla_+\psi_-^a +ims^a \ .\cr}
\eqno (2.8)
$$
Using this result the action (1.2) is readily verified to be invariant under
the
transformations
$$
\eqalign{
\delta_\epsilon\phi^i &= -{i\over2}D_+\epsilon_= \; D_+\phi^i
+\epsilon_=\partial_\dplus\!\phi^i\cr
\Delta_\epsilon\psi_-^a &=-{i\over2} D_+\epsilon_=\;
\nabla_+\psi_-^a +\epsilon_=\nabla_\dplus\!\psi_-^a\cr}
\eqno (2.9)
$$
for $x$-independent (but $\theta$-dependent) superfield parameter $\epsilon_=
\epsilon_=| +i\theta^+\epsilon_-$. The $\epsilon_-$ part of these
transformations can be rewritten as
$$
\delta\phi^i = -{1\over2}\epsilon_- Q_+\phi^i \qquad \delta\psi_-^a =
-{1\over2}\epsilon_-
Q_+\psi_-^a
\eqno (2.10)
$$
where
$$
Q_+ = -iD_+ + 2i\theta^+\partial_\dplus\!\phi^i = \partial_+ +
i\theta^+\partial_\dplus
\eqno (2.11)
$$
is the (hermitian)  differential operator that generates (1,0) supersymmetry
transformations (satisfying $\{Q_+,D_+\} =0$). The $\epsilon_-$ part of the
transformations (2.9) are therefore those of the manifest (1,0) supersymmetry.
The remaining $\epsilon_= |$ part is the transformation generated by the
$P_\dplus$ component of the 2-momentum. By combining the two transformations
they become expressible in terms of (1,0) superfields. The
symmetry transformations generated by the $P_=$ component of the 2-momentum may
similarly be expressed in (1,0) superfield form as
$$
\delta\phi^i =\epsilon_\dplus\partial_=\phi^i \qquad \Delta\psi_-^a =
\epsilon_\dplus\nabla_=\psi_-^a
\eqno (2.12)
$$
where the parameter $\epsilon_\dplus$ is a constant (independent of both $x$
and $\theta^+$).

\vfill\eject
\noindent {\bf 3. (p,0) Supersymmetry }
\medskip

Any additional supersymmetries of (1.2) of the same chirality must have Noether
charges that anticommute with the first one. This implies that the additional
supersymmetry transformations can be expressed in terms of (1,0) superfields
and
a set of constant, anticommuting, parameter(s) $\{\eta^r_- \ ;r=1,\dots
,p-1\}$.
The form of these transformations for $m=0$ is fixed by dimensional analysis;
when $m\ne 0$ we must allow for an additional variation of $\psi$ proportional
to $m$. We are thus led to consider
$$
\eqalign{
\delta_\eta\phi^i &=i\eta^r_- I_{r}{}^i{}_j(\phi)D_+\phi^j\cr
\Delta_\eta\psi_-^a &={1\over2}\eta^r_-{\hat I}_r{}^a{}_b(\phi){\cal S}^b
+{im\over2}\eta^r_- t_r^a(\phi)\cr}
\eqno (3.1)
$$
where $I_r$ are tensors on ${\cal M}$, and $t_r$ and $\hat I_r$  are
sections of $\xi^*$ (the dual of $\xi$) and $\xi\otimes \xi^*$, respectively.
We shall now investigate the conditions for {\sl on-shell} closure of the (p,0)
supersymmetry transformations (i.e. using ${\cal S}_{i-}=0$ and ${\cal
S}_a=0$). We then determine the conditions for invariance of the action. The
constraints arising from these two requirements are necessary and sufficient
for
the existence of conserved charges $\{ Q_+^I\ ; I=1,\dots p\}$ obeying the
(p,0)
supersymmetry algebra.  Off-shell closure of (p,0) supersymmetry algebra
requires stronger conditions, which were investigated in [1].

The conditions for closure of the algebra on $\phi^i$ are
$$
I_rI_s = -\delta_{rs} + f_{rs}{}^t I_t
\eqno (3.2)
$$
(in matrix notation) and
$$
N(I_r,I_s)^i{}_{jk} =0
\eqno (3.3)
$$
where $f_{rs}{}^t$ is zero for p=2 and equal to the quaternion structure
constants $\epsilon_{rst}$ for p=4, and $N$ is the generalised Nijenhuis tensor
defined by
$$
N(I_r,I_s)^i{}_{jk} \equiv 2\big[\partial_l I_r{}^i{}_{[k} I_s{}^l{}_{j]}
- I_r{}^i{}_l \partial_{[j} I_s{}^l{}_{k]} + (r\leftrightarrow s)\big]\ .
\eqno (3.4)
$$
On-shell closure on $\psi_-^a$ implies
$$
F_{ij}{}^a{}_b I_r{}^i{}_{[k} I_s{}^j{}_{l ]} = F_{kl}{}^a{}_b\; \delta_{rs}
\eqno (3.5)
$$
for $m=0$. In addition, if $m\ne 0$ we find the condition
$$
\big[I_r{}^i{}_j \nabla_i G_s^a + (r\leftrightarrow s)\big] +2\delta_{rs}
\nabla_i s^a =0\ .
\eqno (3.6)
$$
There is no condition on $\hat I$ from on-shell closure since $\delta_\eta
{\cal S}^a =-i\eta^r_- \nabla_+(\hat I_r{}^a{}_b{\cal S}^b)$ (when the above
conditions are satisfied) and this vanishes on-shell.

If $m=0$ the action (1.2) is invariant under the transformations (3.1) provided
that
$$
I_r{}^k{}_{(i}g_{j)k} =0 \qquad \nabla_i^{(+)} I_r{}^j{}_k =0
\eqno (3.7)
$$
and
$$
\hat I_{(ab)}\equiv h_{c(a}\hat I^c{}_{b)} =0
\eqno (3.8)
$$
(with no condition on $\hat I_{[ab]}$). In addition, if $m\ne 0$ we find that
$$
\partial_i \big( t_r^as_a \big) =0
\eqno (3.9)
$$
and
$$
\nabla_i t_r^a = I_r{}^j{}_i \nabla_j s^a \ .
\eqno (3.10)
$$

We shall begin the analysis of these conditions by considering the (2,0) case.
The conditions arising from both closure of the supersymmetry algebra and
invariance of the action can be summarised as follows, for $m=0$ [3]: ${\cal
M}$ is a complex manifold with complex structure $I$; $g$ is an Hermitian
metric
with respect to $I$, and the holonomy of the connection $\Gamma^{(+)}$ is a
subgroup of $U(D/2)$. Furthermore (3.5) implies that the vector bundle $\xi
\otimes \com$  is holomorphic\footnote{${}^*$}{Note that, in contrast to
off-shell supersymmetry, on-shell supersymmetry does not require $\hat I$ to
be a complex structure so the rank of $\xi$ is not necessarily even.}. The
additional conditions that arise for $m\ne0$ are just (3.9) and (3.10) since
(3.6) is implied by (3.2) and (3.10). To discuss these conditions it is
convenient to choose complex coordinates $\{\phi^i\}\rightarrow \{\phi^\mu,
\bar\phi^{\bar\mu} \equiv(\phi^\mu)^*\}$ adapted to the complex structure $I$.
Condition (3.10) then reduces to $$ \nabla_{\bar\mu}(s+it)^a=0\ ;
\eqno (3.11)
$$
i.e. $s$ is the real part of a holomorphic section of the bundle $\xi\otimes
{\com}$. The integrability condition of (3.11) is precisely (3.5). Combining
(3.11) with (3.9) we deduce that
$$
s_as^a = t_at^a + {\rm const}.\ \  .
\eqno (3.12)
$$

We turn now to $p=4$. The conditions for $m=0$ can be summarised as follows:
${\cal M}$ admits a quaternionic structure, i.e the three (integrable) complex
structures obey the algebra of imaginary unit quaternions, the metric $g$ is
tri-Hermitian and the holonomy of the connection $\Gamma^{(+)}$ is a subgroup
of $Sp(D/4)$. Furthermore, (3.5) implies that the bundle $\xi\otimes {\com}$ is
holomorphic with respect to all three complex structures. The additional
conditions arising for $m\ne0$ are (3.9) and (3.10) since (3.6) is again
implied by (3.2) and (3.10). The integrability conditions of (3.10) are eqs.
(3.4) and (3.5). It is again convenient to discuss the conditions (3.9) and
(3.10) by choosing complex coordinates adapted to any one of the three complex
structures (it is not possible, in general, to find a coordinate system such
that all complex structures are {\sl simultaneously} constant). Let us choose
coordinates adapted to the complex structure $I_{{}_1}$. From (3.10) we may
then derive the two conditions  $$
\nabla_{\bar\mu}(s+it_{{}_1})^a=0 \qquad \nabla_{\bar\mu}(t_{{}_2}-it_{{}_3})^a
=0\ ,
\eqno (3.13)
$$
i.e. $s+it_{{}_1}$ and $t_{{}_2}-it_{{}_3}$ are holomorphic sections (with
respect to $I_{{}_1}$) of the bundle $\xi\otimes{\com}$. The conditions (3.9)
combined with (3.13) become
$$
s_as^a = h_{ab}t_{{}_1}^a t_{{}_1}^b + {\rm const}.\ \  .
\eqno (3.14)
$$
By adopting coordinates adapted to each of the other two complex structures one
can similarly
deduce the cyclic permutations of the above relations.

\bigskip
\noindent {\bf 4. (p,1) Supersymmetry }
\medskip

Since the (p,1) models are all special cases of (1,1) we shall make use
here of the (1,1) results given in [1]. The vector bundle $\xi$ is
now isomorphic to the tangent bundle, which allows us to convert all bundle
indices to tangent bundle indices. In addition, the connection $\Omega$ of the
bundle $\xi$ becomes the $\Gamma^{(-)}$ connection of the tangent bundle. To
allow for a central charge in the (1,1) supersymmetry algebra, one supposes
that the manifold ${\cal M}$ has an isometry generated by a Killing vector
field $X$, i.e.
$$
\nabla_{(i}X_{j)}=0\ .
\eqno (4.1)
$$
Then, in the presence of torsion, invariance of the action
requires that ${\cal L}_X H=0$ which implies that the two-form $\iota_X H$ is
closed. Thus,
$$
X^k H_{ijk} = \partial_{[i}u_{j]}
\eqno (4.2)
$$
for some locally defined vector $u_i$. In fact, $u$ is globally defined on
${\cal M}$ since the section $s$ of $\xi$ that determines the potential is now
given by
$$
s_i =   u_i - X_i\ .
\eqno (4.3)
$$
The action can now be rewritten as
$$
S=\int\! d^2 x d\theta^+ \big\{
D_+\phi^i\partial_=\phi^j (g_{ij} +b_{ij}) +i\psi_-^i\nabla^{(-)}_+\psi_-^j \,
g_{ij} +im\, s_i\psi_-^i\big\}\ .
\eqno (4.4)
$$
In addition
$$
\partial_i(X^j u_j)=0\ .
\eqno (4.5)
$$
The potential $V$ for the (1,1) model is
$$
V(\phi) = {1\over4} m^2  g^{ij} (u-X)_i(u-X)_j\ .
\eqno (4.6)
$$
Using (4.5)  this can  be rewritten as
$$
V= {1\over4} m^2  (g^{ij}u_iu_j  + g^{ij}X_iX_j ) +\ {\rm const.}\ .
\eqno (4.7)
$$

The action (4.4) is invariant under the (1,0) and (0,1) transformations,
$$
\eqalign{
\delta_\epsilon\phi^i &=-{i\over2}D_+\epsilon_= D_+\phi^i
+\epsilon_=\partial_\dplus\phi^i\cr
\delta_\epsilon\psi_-^i &= -{i\over2}D_+\epsilon_= D_+\psi^i_-
+\epsilon_=\partial_{\dplus}\psi_-^i \ ,\cr}
\eqno (4.8)
$$
and
$$
\eqalign{
\delta_\zeta\phi^i &=D_+\zeta \psi_-^i +m\zeta X^i\cr
\delta_\zeta\psi_-^i &=-iD_+\zeta\partial_=\phi^i +m\zeta\partial_jX^i\psi_-^j\
,}
\eqno (4.9)
$$
respectively [1].

The extended (p,0) transformations take the form
$$
\eqalign{
\delta_\eta\phi^i &=i\eta^r_- I_{r}{}^i{}_j(\phi)D_+\phi^j\cr
\Delta_\eta\psi_-^i &={1\over2}\eta^r_-{\hat I}_r{}^i{}_j(\phi){\cal S}^j
+{im\over2}\eta^r_- t_r^i(\phi)\cr}
\eqno (4.10)
$$
where ${\cal S}^i=0$ is the $\psi_-$ field equation (${\cal S}^i_-=0$ being the
field equation for $\phi^i$) and

For simplicity we will use freely all the conditions above and those derived in
the previous section. The action is therefore invariant and the only commutator
that needs to be checked to ensure (p,1) supersymmetry is the  $\zeta$-$\eta$
one. A calculation yields the following
result
$$
\eqalign{
[\delta_\eta,\delta_\zeta]\phi^i &= -im(\zeta\eta_-^r)D_+\phi^k({\cal
L}_{{}_X}I_r{}^i{}_k) +
im D_+(\zeta\eta_-^r)Z_r^i -{1\over2}(\eta^r_- D_+\zeta)(\hat
I_r{}-I_r)^i{}_j{\cal S}^j\cr
[\delta_\eta,\delta_\zeta]\psi_-^i
&=-[\delta_\eta,\delta_\zeta]\phi^j \Gamma^{(-)}_{jk}{}^i\psi_-^k
+im D_+(\zeta\eta_-^r)V_r{}^i{}_k\psi_-^k
-i{m^2\over2}(\zeta\eta_-^r){\cal
L}_{{}_X}t_r^i\cr & +\big[{1\over2}(D_+\zeta\eta^r_-)\psi_-^k\nabla^{(+)}_k\hat
I_r{}^i{}_l -
{m\over2}(\zeta\eta_-^r){\cal L}_{{}_X}\hat I_r{}^i{}_l\big] {\cal S}^l
-{1\over2}(\eta_-^rD_+\zeta)(\hat I_r -I_r)^i{}_j {\cal S}_-^j }
\eqno (4.11)
$$
where
$$
\eqalign{
Z_r^i &\equiv {1\over2}\big( t_r^i + I_r{}^i{}_j(s^j+2X^j)\big)\cr
V_{r\, ij}&\equiv -\nabla_{[i}t_{r\, j]}\ .}
\eqno (4.12)
$$
The right-hand side of the above commutators is necessarily a symmetry of the
action. The terms which vanish with the field equations leave the action
invariant, so the remaining terms must do so too. Of these terms, note that the
those proportional to the parameter $(\zeta\eta_-^r)$ have the same form as the
(p,0) transformations of (4.10). Here, however, the parameter is {\sl not}
$\theta$-independent so we must impose the conditions
$$
{\cal
L}_{{}_X}I_r{}^i{}_k=0\qquad {\cal L}_{{}_X}t_r^i=0 \ .
\eqno (4.13)
$$

Using (3.7) and (3.10) one can re-express $V_r$ as
$$
V_r^i{}_j= \nabla^{(+)}_j Z_r^i\ .
\eqno (4.14)
$$
The transformations appearing on the right-hand-side of (4.11) that survive
on-shell may now be rewritten as
$$
\eqalign{
\delta_\lambda\phi^i &=  m\lambda^r Z_r^i\cr
\delta_\lambda\psi_-^i &= m\lambda^r\partial_j Z_r^i\psi_-^j\cr}
\eqno (4.15)
$$
where $\lambda^r= i D_+(\zeta) \eta^r_-$. Observe that this takes the
same form as the transformation with parameter $\zeta$, generated by the
Killing
vector $X$, in (4.9). As just remarked, these transformations are
necessarily symmetries of the action and this implies that $Z_r$ are
Killing vector fields (which is also easily seen from the calculation leading
to (4.11)) and leave invariant the sigma-model couplings $H_{ijk}$ and $s$.

We now turn to the calculation of the commutators of the transformations (4.15)
among themselves (for $p=4$) and with the supersymmetry transformations.
The commutators of (4.15) with the $(1,0)$ supersymmetry transformations
vanish. The commutators of (4.15) with the extended $(p,0)$ supersymmetry
transformations are
$$
\eqalign{
[\delta_\lambda,\delta_\eta]\phi^i &= im\eta_-^r\lambda^s D_+\phi^j
\big({\cal L}_{Z_s}I_r{}^i{}_j\big)\cr
[\delta_\lambda,\delta_\eta]\psi_-^i &= {1\over2}m\eta_-^r\lambda^s
\big({\cal L}_{Z_s}\hat I_r{}^i{}_j\big){\cal S}^j +
{i\over2}m^2\eta_-^r \lambda^s [Z_s,t_r]^i -
[\delta_\lambda,\delta_\eta]\phi^j \Gamma^{(-)}_{jk}{}^i \psi_-^k\ .}
\eqno (4.16)
$$
The commutators of (4.15) with the $(0,1)$ supersymmetry transformations
vanish as a consequence of (4.13). Finally, the commutators of the
transformations (4.15) among themselves are
$$
\eqalign{ [\delta_\lambda,\delta_{\lambda'}]\phi^i &= m^2\lambda^s\lambda'{}^r
[Z_s,Z_r]^i\cr
[\delta_\lambda,\delta_{\lambda'}]\psi_-^i &=
m^2\lambda^s\lambda'{}^r\partial_j[Z_s,Z_r]^i\psi_-^j\ .}
\eqno (4.17)
$$
The transformations appearing on the right-hand side of these commutators are
necessarily symmetries of the action. They have the general structure of
$(p,0)$ supersymmetry and (4.15) transformations, respectively. Because of this
the {\sl weakest} condition we can impose is that these new symmetries be
linear
combinations of the existing supersymmetry and (4.15) transformations,
respectively, which will be the case if
$$
\eqalign{
{\cal L}_{Z_s}I_r{}^i{}_j &= A_{sr}{}^t I_t{}^i{}_j \qquad
[Z_s, t_r]^i =  A_{sr}{}^t t_t^i\cr
[Z_s, Z_r]^i &= B_{sr}{}^t Z_t^i}
\eqno (4.18)
$$
where $A$ and $B$ are structure constants, which are restricted by Jacobi
identies.

If $A$ or $B$ is non-zero the supersymmetry algebra is {\sl not} of the form
assumed in the introduction because the additional scalar charges associated
with the invariance under (4.15) are then not central. In principle one
might wish to consider scalar charges that are not central but we leave the
investigatation of this point to the future. Hence we shall require, for {\sl
on-shell} closure, that $A=B=0$, i.e. that
$$
{\cal L}_{Z_s}I_r{}^i{}_j=0\qquad [Z_s, t_r]^i = 0
\qquad [Z_s, Z_r]^i=0\ .
\eqno (4.19)
$$
We remark that the result stated above for the commutators of $(0,1)$
supersymmetry with (4.15) implies that the commutators of the transformations
generated by $X$ commute with those generated by $Z_r$.

We now have all the conditions required for {\sl on-shell} closure of $(p,1)$
transformations, and invariance of the action. Not all of these conditions
are independent. For example, (4.13) and (4.19) are easily seen to be
consequences of the other conditions if $t_r$ is
expressed in terms of the central charge generator $Z_r$ by the first of eqs.
(4.12). Furthermore, substituting the result for $t_r$ into (3.10) we deduce
that
$$
2\nabla_i^{(-)}Z_r^k
-(\nabla_i^{(-)}I_r{}^k{}_j)(s^j+2X^j)-I_r{}^k{}_j(\nabla_i^{(-)}s^j
+2\nabla_i^{(-)}X^j)-I_r{}^j{}_i\nabla_j^{(-)}s^k=0\ .
\eqno (4.20)
$$
Using freely all the conditions derived previously, one finds after some
computation that this equation is equivalent to
$$
2\nabla^{(-)}_{[i}Z_r{}_{j]} -2(u+X)^l H_{ijk} I_r{}^k{}_l +
I_r{}^k{}_j\nabla^{(-)}_k u_i - I_r{}^k{}_i\nabla^{(-)}_k u_j =0\ .
\eqno (4.21)
$$
This is in turn equivalent to
$$
(Z_r+v_r)_i + I_r{}^k{}_i (X+u)_k =0
\eqno (4.22)
$$
where $v_r$ is defined locally by
$$
(Z_r)^k H_{kij} =\partial_{[i} (v_r)_{j]}\ .
\eqno (4.23)
$$
Actually, (4.21) implies only that the one-form defined by the left-hand side
of
(4.22) is closed, so that it can be written locally as the
derivative of a scalar. One then arrives at (4.22) by absorbing this scalar
into the definition of $v_r$.

Note that, by using (4.22) to eliminate $Z_r$ in the first of eqs. (4.12),
$t_r$ can be expressed as
$$
t_r = (Z-v)_r\ .
\eqno (4.24)
$$
Using (3.12) (or its cyclic permutations) we now see that the potential $V$ can
be expressed in terms of $Z_r$ and $v_r$ (for each value of r) in the same way
as it was expressed in terms of $X$ and $u$ in (4.7), i.e.
$$
V= {m^2\over 4} g^{ij}(v_r -Z_r)_i(v_r -Z_r)_j\quad (r=1,\dots,p-1)\ .
\eqno (4.25)
$$
Furthermore, the condition (3.9) can now be re-expressed as
$$
u\cdot v_r + X\cdot Z_r = {\rm const.}
\eqno (4.26)
$$
but this can be shown to be a consequence of the other conditions including, in
particular, (4.22). We shall not present the details of the calculation here
since a very similar calculation will be described in the next section for the
(1,q) models. Analogous calculations lead also to the relations
$$
v_r\cdot v_s + Z_r\cdot Z_s = {\rm const.}\qquad (r\ne s)
\eqno (4.27)
$$
A consequence of the relations (4.26) and (4.27) is that the potential $V$ can
be rewritten as
$$
V= {m^2\over 4|c|^2}g^{ij}\big( c\cdot [{\bb X}-{\bbbb U}]\big)_i \big( c\cdot
[{\bb X}-{\bbbb U}] \big)_j
\eqno (4.28)
$$
where $[{\bb X}-{\bbbb U}]$ is a p-vector in the space of central charges with
components $\big(X-u, Z_r -v_r\big)$ and $c$ is {\sl any} constant
p-vector in this space. Despite the explicit appearance in the potential of
$c$, the potential is $O(p)$ invariant because it is actually {\sl independent}
of the choice of $c$ (up to a constant).

This result agrees with that obtained in [1] for (1,1) supersymmetry on the
assumption of {\sl off-shell} closure of the supersymmetry algebra. Off-shell
closure requires $\hat I_r=I_r$, but  here, in contrast to the (p,0) case, this
places no conditions on the {\sl action} that were not already required for
on-shell closure, so we may choose $\hat I_r=I_r$ without loss of generality.

\medskip
{\bf 5. (1,q) supersymmetry}
\medskip
We now investigate the conditions under which the action (4.4) has (1,q)
supersymmetry. The results are of course equivalent to those just obtained for
(p,1), again assuming that any scalar charges appearing in the supersymmetry
commutators are central. By subsequently combining the (1,q) with the (p,1)
results we shall be able to discuss the remaining (p,q) models.

On dimensional grounds the extended (0,q) supersymmetry transformations can be
written, in
terms of (1,0) superfields, as
$$
\eqalign{ \delta_{\kappa^r}\phi^i &= D_+\kappa^r J_r{}^i{}_j
\psi_-^j +m\kappa^r Y_r^i(\phi)\cr
\Delta_{\kappa^r}\psi_-^i &= -iD_+\kappa^r
\hat J_r{}^i{}_j\partial_=\phi^j +D_+\kappa^r
L^r{}^i{}_{jk}(\phi)\psi_-^j\psi_-^k +
m\kappa^r W_r{}^i{}_j\psi_-^j\cr}
\eqno (5.1)
$$
where $J$, $\hat J$, $L$, $Y$ and $W$ are tensors on ${\cal M}$, and
$\kappa^r(\theta)$ are $\theta$-dependent but $x$-independent (1,0) superfield
parameters.

Using freely the conditions obtained previously from (1,1) supersymmetry, the
following conditions result from requiring closure of the algebra of the (0,q)
extended supersymmetry transformations (5.1) when $m=0$:
$$
\hat J_r=-J_r\ ,
\eqno (5.2)
$$
$$
L_r^i{}_{jk}= \nabla^{(-)}_{[k}J_r{}^i_{j]} + H_{jk}{}^m J_r^i{}_m +
2H_{l[j}{}^i
J_r{}^l{}_{k]}
\eqno (5.3)
$$
$$
J_rJ_s =-\delta_{rs} + f_{rs}{}^t J_t\ ,
\eqno (5.3)
$$
and the Nijenhuis conditions
$$
N(J_r,J_s)=0\ .
\eqno (5.4)
$$
For $m\ne0$ we find the following additional conditions:
$$
W_r{}^i{}_j =\nabla^{(+)}_j Y_r^i
\qquad[X,Y_r] =0\qquad [Y_r,Y_s]=0
\eqno (5.5)
$$
and
$$
{\cal L}_{{}_X} J_r=0 \qquad {\cal L}_{{}_Y} J_r =0 \ .
\eqno (5.6)
$$
We have still to check closure of the algebra of the (0,q) with the (1,0)
supersymmetry transformations, but this we leave for the moment.

The invariance of the action under the  (0,q) supersymmetry transformations
imposes various new conditions. Simplifying these with the aid of those derived
above, we find (after some computation) the following {\sl independent}
additional conditions. Firstly, for $m=0$,
$$
J_{r\, (ij)}=0 \qquad \nabla^{(-)}_iJ_r^j{}_k =0\ .
\eqno (5.7)
$$
For $m\ne0$ the new independent conditions are
$$
\nabla_{(i}Y_{r\, j)}=0 \qquad W_r{}_{ij}= J_r^k{}_i J_r^l{}_jW_r{}_{kl}
\eqno (5.8)
$$
i.e. that the vector fields $Y_r$ are Killing and the tensor $W_r$ is (1,1)
with respect to the complex structure $J_r$. We further find that the Lie
derivative of the torsion with respect to $Y_r$ must vanish, a condition that
is
(locally) equivalent to
$$
Y_r^iH_{ijk}=\partial_{[j}w_{r\, k]}\ ,
\eqno (5.9)
$$
which defines $w_r$ up to the gradient of a locally defined function. Also, the
section $s$
satisfies
$$
J_r{}^i{}_j\, s^j   = (w_r- Y_r)^i
\eqno (5.10)
$$
and ({\it no summation over the index r})
$$
\partial_i(Y_r^j w_{r\, j})=0\quad (r=1,\dots, p-1)\ .
\eqno (5.11)
$$

We next compute the commutator of the (0,q) with the (1,0) supersymmetry
transformations. Taking into account that the parameters are superfields and
hence $\theta$-dependent, we
find that
$$
\eqalign{
[\delta_\kappa,\delta_\epsilon]\phi^i &= -{i\over2} m D_+\epsilon_= D_+\kappa^r
Y_r^i\cr
[\delta_\kappa,\delta_\epsilon]\psi_-^i &= -{i\over2}m D_+\epsilon_=
D_+\kappa^r
W_r{}^i{}_j\psi_-^j
-[\delta_\kappa,\delta_\epsilon]\phi^k\Gamma^{(-)}_{kj}{}^i\psi_-^j\cr}
\eqno (5.12)
$$
i.e. that when $m\ne0$ there are $p-1$ possible additional central charges
associated with the Killing vector fields $Y_r$. Clearly the $Y_r$ are the
analogues of the Killing vector fields $Z_r$ found in the previous section.
Finally, it can be verified that the transformations generated by the
vector fields $Y_r$ are indeed those of {\sl central} charges, without the
need for imposing any further conditions.

We turn now to the form of the potential $V$. Following roughly the same steps
as for the (p,1) case one can show that the potential $V$ can
be expressed in terms of $Y_r$ and $w_r$ (for each value of r) in the same way
as it was expressed in terms of $X$ and $u$ in (4.7), i.e.
$$
V= {m^2\over4} g^{ij}(w_r -Y_r)_i(w_r -Y_r)_j\quad (r=1,\dots,p-1)\ .
\eqno (5.13)
$$
Note now that by contracting (5.10) with the vector $(u-X)$ one has
$$
(u\cdot w_r + X\cdot Y_r) = (u\cdot Y_r + X\cdot w_r)\ .
\eqno (5.14)
$$
But the right-hand-side of (5.14) is constant. This can be seen as follows:
$$
\eqalign{
d(u\cdot Y_r + X\cdot w_r)&= d\iota_{Y_r} u + d\iota_X w_r\cr
&= -(\iota_{Y_r} du + \iota_X dw_r)\cr
&= -(\iota_{Y_r}\iota_X H + \iota_X\iota_{Y_r} H)\equiv 0\ .}
\eqno (5.15)
$$
One can show by a similar argument that
$$
w_r\cdot w_s + Y_r\cdot Y_s ={\rm const.}\qquad (r\ne s)
\eqno (5.16)
$$
Following the same reasoning as in the (p,1) case the potential $V$ can be
expressed in the form
$$
V= {m^2\over 4|c|^2}g^{ij}\big( c\cdot [{\bb X}-{\bbbb U}]\big)_i \big( c\cdot
[{\bb X}-{\bbbb U}] \big)_j
\eqno (5.17)
$$
where $[{\bb X}-{\bbbb U}]$ is a now a q-vector in the space of central charges
with components $\big(X-u, (Y_r -w_r)\big)$ and $c$ is again {\sl any}
constant p-vector in this space. As expected the final results for (1,q) are
equivalent to those found previously for (p,1) supersymmetry.

\bigskip
\noindent {\bf 6. (p,q) Supersymmetry}
\medskip

We are now in a position to investigate the conditions under which the action
(4.4) has (p,q) supersymetry for p and q {\sl both} greater than 1. The
conditions for invariance of the action are just those obtained previously for
either (p,1) or (1,q) supersymmetry. The only additional requirement for (p,q)
supersymmetry is the closure of the algebra of the extended (p,0) with the
extended (0,q) transformations. Since these correspond to supersymmetry charges
of opposite chirality we expect additional central charges. Using freely
conditions previously derived, a calculation of the commutator on $\phi$ leads
to the following result:
$$
[\delta_\eta,\delta_\kappa]\phi^i =im D_+(\kappa^s \eta_-^r) Z_{sr}
-im\kappa^s\eta_-^r D_+\phi^k({\cal L}_{Y_s} I_r{}^i{}_k)
+{1\over2}D_+(\kappa^s\eta_-^r)
(J_s\hat I_r - I_r J_s)^i{}_j {\cal S}^j
\eqno (6.1)
$$
where
$$
Z_{sr} \equiv {1\over2}\big( J_s{}^i{}_j t_r^j + I_r{}^i{}_j (2Y_s^j +
J_s{}^j{}_k s^k) \big)
\eqno (6.2)
$$
A similar calculation for $\psi$ leads to
$$
\eqalign{
[\delta_\eta,\delta_\kappa]\psi_-^i =& im
D_+(\kappa^s\eta_-^r)V_{sr}{}^i{}_j\psi_-^j
-[\delta_\eta,\delta_\kappa]\phi^k\Gamma^{(-)}_{kj}{}^i \psi_-^j\cr
&+{1\over2}D_+(\kappa^s\eta_-^r)(J_sI_r -\hat I_r J_s)^i{}_k {\cal S}_-^k
-{1\over2}m\kappa^s\eta_-^r({\cal L}_{Y_s}\hat I_r{}^i{}_l){\cal S}^l\cr
&+D_+(\kappa^s\eta_-^r)\psi_-^m\big[{1\over2}J_s{}^k{}_m\nabla^{(+)}_k\hat
I_r^{il} + 2\hat
I_r^{k[i} L_s^{l]}{}_{mk} +2\hat I_r^{k[i} H^{l]}{}_{qk}
J_s{}^q{}_m\big]{\cal S}_l\cr &-{i\over2}m^2\eta_-^r\kappa^s[Y_s,t_r]^i\cr}
\eqno (6.3)
$$
where
$$
V_{sr}{}^i{}_j \equiv -\big[ L_s{}^i{}_{jk}t_r^k
+{1\over2}(J_sI_r)^{ik}\nabla^{(-)}_ks_j
-{1\over2}J_s{}^m{}_j\nabla^{(-)}_mt_r^i\big]\ .
\eqno (6.4)
$$
The terms that vanish with the field equations are (trivially) symmetries of
the action so
the remaining terms must also be symmetries. Of these, the terms with
coefficient
$\kappa^s\eta_-^r$ have the form of an $\eta$-supersymmetry, but the
$\theta$-dependence of
this coefficient precludes this identification and we must set
$$
{\cal L}_{Y_s} I_r =0\qquad [Y_s,t_r]=0\ .
\eqno (6.5)
$$
The remaining (on-shell) transformations may now be identified as those of new
central charge symmetries. In particular, it follows that the vector fields
$Z_{sr}$ are Killing, that the torsion and the section $s$ are invariant
with respect to these symmetries, and that $V_{sr}{}^i{}_j
=\nabla^{(+)}_jZ_{sr}^i$, exactly as for the previous central charges. The new
central charge transformations may be simplified to
$$
\eqalign{ \delta_\lambda\phi^i &=  m\lambda^{sr} Z_{sr}^i\cr
\delta_\lambda\psi_-^i &= m\lambda^{sr}\partial_j Z_{sr}^i\psi_-^j\ .}
\eqno (6.6)
$$
where $\lambda_{sr}$ is a constant parameter.
Finally, the commutators of the new central charge
transformations with themselves and with with all other central charge and
supersymmetry transformations will vanish, as required, provided that all
commutators of the Killing vector fields $Z_{sr}$ with themselves and with the
other Killing vector fields vanish, and provided that the Lie derivatives
of the complex structures $I_r$ and $J_s$ with respect to $Z_{sr}$ vanish.

As in the (p,1) case we may now use (6.2) and (3.10) to
eliminate $t_r$. Following the same steps as those described in previous
sections, one finds that
$$
 (Z_{sr} +v_{sr})_i + I_r{}^k{}_i (Y_s + w_s)_k =0\ ,
\eqno (6.7)
$$
where $v_{sr}$ is locally defined by $\iota_{Z_{sr}}H= dv_{sr}$. Using this
result in (6.2) we now find that
$$
J_s{}^i{}_j t_r^j = -(v_{sr} - Z_{sr})^i\ .
\eqno (6.8)
$$
This implies that $V$ can be written in pq different ways as
$$
V= {m^2\over 4} g^{ij}(v_{sr} -Z_{sr})_i(v_{sr} -Z_{sr})_j\quad
\pmatrix{r=1,\dots,p-1\cr s= 1\dots ,q-1\cr}\ .
\eqno (6.9)
$$

In analogy with (5.17), the potential can now be rewritten, up to a constant,
as
$$
V= {m^2\over 4|c_s|^2}g^{ij}\big( c_s\cdot [{\bb Y}_s -{\bbbb W}_s]\big)_i
\big( c_s\cdot [{\bb Y}_s -{\bbbb W}_s]\big)_j\qquad (s=1,\dots q-1)\ .
\eqno (6.10)
$$
where $[{\bb Y}_s -{\bbbb W}_s]$ is a p-vector (for each value of $s$) with
components $\big(Y_s- w_s, Z_{sr}- v_{sr}\big)$, and $c_s$ is a
p-vector (for each value of $s$) with constant components. To establish this
result one notices that the cross terms sum to a constant as a result of (6.7)
while the diagonal terms are all equal. The details follow the same lines as
in the previous two sections.

Similarly, from (4.24) and (6.8) one can deduce that the potential can also be
written as
$$
V= {m^2\over 4|c_r|^2}g^{ij}\big( c_r\cdot [{\bb Z}_r -{\bbbb V}_r]\big)_i
\big( c_r\cdot [{\bb Z}_r -{\bbbb V}_r]\big)_j\qquad (r=1,\dots p-1)
\eqno (6.11)
$$
where $[{\bb Z}_r -{\bbbb V}_r]$ is a q-vector (for each value of $r$) with
components $\big(Z_r- v_r, Z_{sr}- v_{sr}\big)$, and $c_r$
is a q-vector (for each value of $r$) with constant components.

This concludes our discussion of the conditions required for {\sl on-shell}
(p,q) supersymmetry. In subsequent sections we shall consider their
implications
for certain interesting special cases. We turn now to the conditions required
for {\sl off-shell} closure of the supersymmetry algebra. From our discussion
of
the conditions for off-shell (p,1) supersymmetry in section 4, we know that
this requires us to set $\hat I_r =I_r$. From (6.1) we deduce that
the additional requirement for off-shell (p,q) supersymmetry is that, in matrix
notation,
$$
[I_s,J_r]=0
\eqno (6.12)
$$
In fact, this condition is sufficient to show that all field-equation terms in
the commutator (6.3), on $\psi$, vanish. Thus (6.12) is the only additional
condition for off-shell closure of the extended (p,0) and (0,q)
supersymmetries.

\bigskip
\noindent {\bf 7. Potentials for (2,2) and (4,4) models}
\medskip

The supersymmetry algebra of a massive sigma-model with (p,q) supersymmetry has
a possible pq central charges, $Z^{II'}$ in the notation of the
introduction. These correspond to the Killing vector fields $X$, $Y_r$, $Z_s$,
and $Z_{sr}$. We have seen previously that the potential $V$ is expressed in
terms of these Killing vector fields and the associated one-forms $u$, $w_r$,
$v_s$ and $v_{sr}$, respectively. We have seen previously \big(eqs.
(4.22), (5.12), (6.7) and (6.8)\big) that these quantities are constrained by
the relations
$$
\eqalign{
(Z_r + v_r)_i + I_r{}^k{}_i (X+u)_k &= 0\cr
(Y_s-w_s)_i +  J_s{}^k{}_i(X-u)_k &=0\cr
(Z_{sr} + v_{sr})_i + I_r{}^k{}_i (Y_s + w_s)_k &=0\cr
(Z_{sr} + v_{sr})_i + J_s{}^k{}_i (Z_r - v_r)_k &=0}
\eqno (7.1)
$$
where $r= (1,\dots, p-1)$ and $s=(1,\dots, q-1)$.
To complete the determination of the general form of the potential we must
therefore solve these relations for, e.g., $X$ and $u$. This task is greatly
simplified by the observation that the massless model has an $SO(p)\times
SO(q)$ symmetry which translates into an $SO(p)\times SO(q)$ isometry group of
the supersymmety algebra. By means of such an $SO(p)\times SO(q)$
transformation the number of non-zero central charges of the massive model can
be reduced. For example, if $p=q$ a basis of the supersymmetry
charges may always be found for which $Z^{II'}$ is diagonal, i.e. the only
non-zero central charges are those generated by $X$ and $T_r\equiv Z_{rr}$.
This observation would not be so useful if the potential $V$ were not also
invariant under $SO(p)\times SO(q)$ because the form of the potential in a
special basis would then be a special form and our intention is to find the
general form. Fortunately, we showed in the previous section that
the potential $V$ is {\sl also} $SO(p)\times SO(q)$ invariant. Hence no
generality is lost if, for the p=q models with torsion, we set
$$
Y_r=Z_r=0\qquad Z_{rs}=0\  (r\ne s)\ ,
\eqno (7.2)
$$
so that the only non-zero Killing vector fields are $X$ and $T_r\equiv Z_{rr}$.
In this case,
$$
w_r = d c_r \qquad v_s = d b_s \qquad v_{sr} = d e_{sr}\ (r\ne s)
\eqno (7.3)
$$
for locally-defined scalar functions $c_r$, $b_s$ and $e_{sr}$. Substituting
(7.2) and (7.3) into (7.1) we obtain the new relations
$$
\eqalign{
\partial_i b_r + I_r{}^k{}_i (X+u)_k &=0\cr
\partial_i c_s - J_s{}^k{}_i (X-u)_k &=0\ ,}
\eqno (7.4)
$$
$$
\eqalign{
(T_r + n_r)_i + I_r{}^k{}_i\partial_k c_r &=0\cr
(T_s - n_s)_i - J_s{}^k{}_i\partial_k b_s &=0}
\eqno (7.5)
$$
where $n_r\equiv v_{rr}$, and
$$
\eqalign{
\partial_i e_{sr} + I_r{}^k{}_i\partial_k c_s &=0\ (r\ne s)\cr
\partial_i e_{sr} + J_s{}^k{}_i \partial_k b_r &=0\ (r\ne s)\  .}
\eqno (7.6)
$$
Using (7.4), the eqs. (7.5) can be solved for $T_r$ and $n_r$ in terms of $X$
and $u$ as follows
$$
\eqalign{
(T_r)_i &= {1\over2}\{ I_r,J_r\}^k{}_i X_k +{1\over2}[I_r,J_r]^k{}_i u_k\cr
(n_r)_i &= -{1\over2}\{ I_r,J_r\}^k{}_i u_k -{1\over2}[I_r,J_r]^k{}_iX_k \ .}
\eqno (7.7)
$$
Eliminating the functions $e_{sr}$ from (7.6) and then using (7.4) we find the
following constraint on $X$ and $u$:
$$
(u +X)_i = (J_sI_rJ_sI_r)^k{}_i (u-X)_k \ (r\ne s)\ .
\eqno (7.8)
$$
Since the potential $V$ can be expressed entirely in terms of $X$ and $u$,
the relevant relations are those of (7.4) and (7.8). We now turn to a
discussion of the consequences of these relations for the (2,2) and (4,4)
models.

For (2,2) models (7.8) does not apply since, necessarily, $r=s=1$ and
(7.4) is equivalent to
$$
\eqalign{
(X+u)_i &= I^k{}_i\partial_k b\cr
(X-u)_i &= -J^k{}_i\partial_k c \ .}
\eqno (7.9)
$$
These equations generalize the expression in [8] for a holomorphic Killing
vector field in terms of a Killing potential to the case of non-zero torsion.
Note that $V$ is expressed in terms of the $(X-u)$ so the potential is the
square of the derivative of $c$.

Consider now the special case of zero torsion
and $I=J$ discussed in [8]. Solving (7.4) for $u$ and $X$ we get
$$
X_i = -I^k{}_i \partial_k \big({c-b\over2}\big)\qquad
u_i = I^k{}_i \partial_k \big({c+b\over2}\big)\ .
\eqno (7.10)
$$
{}From the first of these equations we identify $(c-b)/2$ as the Killing
potential of the holomorphic Killing vector field $X$. Moreover, since the
torsion vanishes, $u=da$ for some locally defined scalar function $a$. Thus
the second of the equations (7.9) implies that $a$ is the real part of a
locally defined holomorphic function (which is the superpotential in the
superspace formulation). Note further that in this special case $T=-X$. These
results agree with those of [8] but the potential given there was expressed in
terms of {\sl two} commuting Killing vector fields. However, as we have seen,
the general scalar potential can be written (up to a constant) in terms of a
single holomorphic Killing vector field (which must be a linear combination of
those considered in [8]).

We now turn to some special classes of (4,4) models. For (4,4) models eq.
(7.8) is applicable and has important consequences for the potential $V$.
First
consider the zero torsion case, for which $u=da$, and $I_r=J_r$. In this case,
(7.8) implies that $da=0$, and the relation (7.4) allows the identification
of the functions $(c_r-b_r)/2$ with the Killing potentials of the
tri-holomorphic Killing vector field $X$. Note also that $T_r=-X$. We
conclude that the general scalar potential $V$ for (4,4) models with zero
torsion and $I_r=J_r$ is given by the length of a tri-holomorphic
Killing vector field. This is consistent with the results of [6] where such a
model was constructed\footnote{${}^*$}{We have not understand the results of
the
authors of [8] on (4,4) models well enough to know whether their results agree
with ours.}.

Another class of (4,4) models are those for which the supersymmetry algebra
closes off-shell, i.e. $[I_r,J_s]=0$. Note that this is not possible when
$I_r=J_r$. In this case (7.8) and (7.5) imply that $X=0$ and $T_r=0$. Because
$X=0$, $u=da$. The (4,4) analogue of (7.9) now implies that $a$ can be
written in three different ways as the real part of a holomorphic function
with respect to each of the complex structures. Because there are no central
charges in this case, the (4,4)-superfield formalism of [4] applies, so there
exists an off-shell (4,4) superfield action.

\bigskip
\centerline{\bf Acknowledgements}
\medskip
G.P. thanks the Commission of European Communities for financial support.

\bigskip
\centerline{\bf REFERENCES}
\medskip

\item {[1]}
C.M. Hull, G. Papadopoulos and P.K. Townsend, {\it Potentials for (p,0) and
(1,1) supersymmetric sigma-models with torsion}, preprint DAMTP/R-93/8.

\item {[2]}
S.J. Gates, C.M. Hull, M. Ro{\v c}ek, Nucl. Phys. {\bf B248} (1984) 157;
C.M. Hull, in {\it Super Field Theories}, eds. H. Lee and G. Kunstatter
(Plenum, N.Y. 1986).

\item {[3]}
C.M. Hull and E. Witten, Phys. Lett. {\bf 160B} (1985) 398.

\item {[4]}
P.S. Howe and G. Papadopoulos, Class. Quantum Grav. {\bf 5} (1988) 1647; Nucl.
Phys. {\bf B289} (1987) 264.

\item {[5]}
W. Lerche and N.P. Warner, Nucl. Phys. {\bf B358} (1991) 571.

\item {[6]}
E.R.C. Abraham and P.K. Townsend, Phys. Lett. {\bf 291B} (1992) 85; Phys. Lett.
{\bf B295} (1992) 2545.

\item {[7]}
E. Witten, {\it Phases of N=2 Theories in Two Dimensions}, preprint
IASSNS-HEP-93/3.

\item {[8]}
 L. Alvarez-Gaum{\' e} and D.Z. Freedman, Commun. Math. Phys.
{\bf 91} (1983), 87.

\item {[9]}
C.M. Hull, to appear.

\end